\documentclass[preprint1]{aastex6}
\usepackage{amsmath}
\slugcomment{}

\def\eps{\epsilon}
\newcommand{\be}{\begin{equation}}
\newcommand{\ee}{\end{equation}}

\shorttitle{Progenitors of High-$z$ Quasars}
\shortauthors{Fatuzzo \& Melia}

\begin{document}

\title{Unseen Progenitors of Luminous High-$z$ Quasars in the
$R_{\rm h}=ct$ Universe}
\author{Marco Fatuzzo}\affil{Physics Department, Xavier University, Cincinnati, OH 45207; \\
fatuzzo@xavier.edu.}
\author{Fulvio Melia\thanks{John Woodruff Simpson Fellow.}}\affil{Department of Physics, The Applied Math 
Program, and Department of Astronomy, The University of Arizona, AZ 85721, USA \\
fmelia@email.arizona.edu.}

\begin{abstract}
Quasars at high redshift provide direct information on the mass growth of supermassive black holes
and, in turn, yield important clues about how the Universe evolved since the first (Pop III) stars
started forming. Yet even basic questions regarding the seeds of these objects and their growth
mechanism remain unanswered. The anticipated launch of eROSITA and ATHENA is expected to facilitate
observations of high-redshift quasars needed to resolve these issues. In this paper, we compare
accretion-based supermassive black hole growth in the concordance $\Lambda$CDM model with
that in the alternative Friedmann-Robertson Walker cosmology known as the $R_{\rm h} = ct$
universe. Previous work has shown that the timeline predicted by the latter
can account for the origin and growth of the $\ga 10^9\;M_\odot$ highest redshift quasars
better than that of the standard model. Here, we significantly advance this comparison by
determining the soft X-ray flux that would be observed for Eddington-limited accretion
growth as a function of redshift in both cosmologies. Our results indicate that a clear
difference emerges between the two in terms of the number of detectable quasars at redshift
$z \ga 7$, raising the expectation that the next decade will provide the observational
data needed to discriminate between these two models based on the number of detected
high-redshift quasar progenitors. For example, while the upcoming ATHENA mission is 
expected to detect $\sim 0.16$ (i.e., essentially zero) quasars at $z\sim 7$ in $R_{\rm h}=ct$, 
it should detect $\sim 160$ in $\Lambda$CDM---a quantitatively compelling difference.
\end{abstract}

\keywords{}
\section{Introduction}
The ongoing detection of quasars at high ($z\ga 5$) redshift provides vital information regarding
the growth of black hole (BH) mass and, in turn, informs our understanding of how the Universe has
evolved since the beginning of the stelliferous era. To be sure, the assembly of the high-$z$
quasar sample has been painstaking work as these objects have proven to be quite elusive.
As an example, Weigel et al. (2015) estimated that a search for $z\ga 5$ AGNs in the Chandra
Deep Field South (0.03 deg$^2$ field of view) should have lead to a discovery of $\sim 20$ AGNs,
yet no convincing identifications were made.  Such non-detections, while consistent with the seeming
strong evolution at the faint-end of the AGN luminosity function with increasing redshift from $z \sim 3$ to
$z\sim 5$ (Georgakakis et al. 2015), do put existing models of early black hole evolution at odds
with observational constraints on their growth rate (Treister et al. 2013).  Several possible
explanations have been proposed for the very limited number of detections, including dust obscuration
(Fiore et al. 2009), low BH occupation fraction, super-Eddington accretion episodes with low duty cycles
(Madua et al. 2014; Volonteri \& Silk 2015), and BH merging scenarios.

At the same time, the handful of $z > 6.5$ quasars that have been observed (e.g., Mortlock et al. 2011)
provide super-massive black hole (SMBH) mass estimates that are hard to reconcile with the
timeline of a $\Lambda$CDM Universe (Melia 2013; Melia \& McClintock 2015). Specifically,
black holes in the local Universe are produced via supernova explosions with masses
$\approx 5 - 20 \;M_\odot$.  But Eddington-limited accretion would require
$\sim 10^5\;M_\odot$ seeds in order to produce the billion solar-mass quasars seen at redshift
$z\ga 6.5$.  Accretion scenarios operating within the $\Lambda$CDM paradigm thus require
either anomalously high accretion rates (Volonteri \& Rees 2005) or the creation of
massive seeds (Yoo \& Miralda-Escude\'e 2014), neither of which has actually ever been observed.

In recent work, Melia (2013) and Melia \& McClintock (2017) present a simple and elegant solution
to the supermassive black hole anomaly by viewing the evolution of SMBHs through the age-redshift
relation predicted by the $R_{\rm h}=ct$ universe, a Friedmann-Robertson-Walker (FRW) cosmology
with zero active mass. In their scenario, cosmic re-ionization lasted from $t \approx 883$
Myr ($z\sim 15$) to $t \approx 2$ Gyr ($z\sim 6$) (see also Melia \& Fatuzzo 2016).
As such, $5 - 20\;M_\odot$ black hole seeds that formed shortly after the beginning of
re-ionization would have evolved into $\sim 10^{10}\;M_\odot$ quasars by $z\sim 6 - 7$
via the standard Eddington-limited accretion rate.  It should be noted that these SMBH
results are but one of many comparative tests completed between the $R_{\rm h}=ct$
and $\Lambda$CDM paradigms, the results of which show that the data tend to favor the
former over the latter with a likelihood $\sim 90\%$ versus $\sim 10\%$, according to the
Akaike (AIC) and Bayesian (BIC) Information Criteria (see, e.g., Wei et al. 2013; Melia \&
Maier 2013; Melia 2014; Wei et al. 2014a, 2014b; Wei et al. 2015a, 2015b; Melia et al. 2015).
A summary of 18 such tests may be found in Table~I of Melia (2017).

Clearly, our understanding of SMBH evolution remains an open question, with even the basic
questions on how these objects were seeded and the mechanism through which they evolved
remaining unanswered. But the next decade  promises to be transformative. The eROSITA
mission, scheduled for launch in 2018, will perform the first imaging all-sky survey
in the medium energy X-ray range with unprecedented spectral and angular resolution.
Likewise, the ATHENA X-ray observatory mission scheduled for launch in 2028 is expected
to perform a complete census of black hole growth in the Universe tracing to the earliest
cosmic epochs.

Motivated by the feasibility of testing the $R_{\rm h}=ct$  and $\Lambda$CDM paradigms
with this upcoming wealth of observational data, we here extend the analysis of Melia (2013)
and Melia \& McClintock (2017) by using accretion-based evolutionary models of SMBHs
to determine the expected soft X-ray flux values observable at earth as a function of
redshift. The analysis is carried out for both {\it Planck} $\Lambda$CDM, with 
optimized parameters $\Omega_{\rm b} = 0.308$, $w = -1$ and $H_0 = 67.8$ km s$^{-1}$ 
Mpc$^{-1}$, and the $R_{\rm h} = ct$ universe with the same Hubble constant for ease 
of comparison. We shall demonstrate that
the quasar mass function for SMBHs at $z \ga 7$ is considerably different between the
two scenarios, indicating that the next generation of quasar observations at high-redshifts
will allow us to discriminate between these two cosmologies.

The paper is organized as follows. We present our SMBH mass evolutionary model in
\S2, and relate it to redshift evolution in both $\Lambda$CDM and $R_{\rm h} = ct$.
We then present our emission model in \S 3, which links the mass of a black hole to its
X-ray emissivity. In \S4, we combine the results of \S2 and \S3 to calculate the flux
expected as a function of redshift during the evolutionary history of a SMBH for both
the $\Lambda$CDM and $R_{\rm h} = ct$ cosmologies. These results are then combined with
the known quasar mass function at $z = 6$ in order to calculate the expected number
of observable quasars as a function of redshift, again for both cosmologies. Our summary
and conclusions are presented in \S 5.

\section{Steady Eddington-limited Black-Hole Evolution in the Early Universe}
We adopt a streamlined model wherein the early Universe  ($ 6\lesssim z \lesssim 10$) 
is comprised of non-rotating black holes of mass
$M_{\rm bh}$ that grow continuously through mass accretion via a thin or slim disk (see below),
maintaining a constant radiative efficiency $\eps_r$ over that time (see, e.g., Chan et al. 2009).
The ensuing bolometric luminosity is parametrized
in terms of the Eddington ratio  $\lambda_{\rm Edd} \equiv L_{bol} / L_{\rm Edd}$, which is also assumed to
remain constant throughout this time.
The disk accretion rate is therefore given by $\dot M = L_{bol} /(\eps_r c^2)$, and with the further
assumption that a black hole steadily accretes a fraction $(1-\eps_r)$ of the
infalling material (see, e.g., Ruffert \& Melia 1994), the mass growth rate is
\be
\dot M_{\rm bh}  ={(1 - \eps_r) \over \eps_r} { \lambda_{\rm Edd}  L_{\rm Edd} \over c^2} = {(1 - \eps_r) \over \eps_r} { \lambda_{\rm Edd} \over t_{\rm Edd} }\,M_{\rm bh} \,,
\ee
where $t_{\rm Edd} = 0.45$ Gyrs.
Integrating Equation~(1), one obtains an expression for the black hole mass as a function
of the age of the Universe,
\be
M_{\rm bh} (t) = M_0\, e^{{(1-\eps_r)\over\eps_r} { \lambda_{\rm Edd} \over t_{\rm Edd}} (t - t_0)}\,,
\ee
where $M_0$ is the mass observed at redshift $z_0$, corresponding to an age $t_0$.

Connecting Equation~(2) to observational cosmology requires a relation between redshift and the age
of the Universe.  In $\Lambda$CDM, this relation is given via the well known integral expression
\be
t^{\Lambda}(z) = {1\over H_0} \int_z^\infty {du\over\sqrt{\Omega_m (1+u)^5
+ \Omega_\Lambda (1+u)^{5+3w}}}\;,
\ee
where the radiation contribution has been omitted given the redshift of interest, thus leaving
 $\Omega_\Lambda = 1-\Omega_m$.
 The corresponding expression for $R_{\rm h}=ct$ takes on the simpler form
\be
t^{R_{\rm h}}(z) = {1\over H_0 (1+z)}\;.
\ee
We adopt the {\it Planck} parameters (Ade et al. 2016), $H_0 = 67.8$ km s$^{-1}$ Mpc$^{-1}$,
$\Omega_m = 0.308$, and $w= -1$, throughout this work.

The difference in age between $z\sim 10$ and $z\sim 6$ for the two scenarios, as illustrated in figure~1, has
important implications for quasar evolution and detectability, and can therefore be used to test
each model against present and future observations.   Specifically, while the age of the Universe at $z = 6$ is
approximately $0.93$ Gyrs in standard ($\Lambda$CDM) cosmology, its value is approximately 2.1 Gyrs for
$R_{\rm h} = ct$.  As a result, a supermassive black hole in the $R_{\rm h} = ct$ universe
has $\sim 18 t_{\rm Edd}$ of additional time to grow before redshift $z = 6$, and is therefore advantaged
by a factor $\sim e^{18 } \approx 6 \times 10^7$ over its $\Lambda$CDM counterpart (assuming
$\eps_r = 0.1$ and $\lambda_{\rm Edd} = 1$).  More relevant to our discussion, the time
interval between $z = 6$ and $z = 10$ in the $R_{\rm h} = ct$ universe is approximately
$0.74$ Gyrs versus $0.46$ Gyrs in $\Lambda$CDM. As shown in figure~2, mass growth during
this epoch in $R_{\rm h} = ct$ is therefore advantaged by a factor of
$\sim e^{6} \approx 400$ over its $\Lambda$CDM counterpart for this scenario.
These dramatically different growths, as measured by redshift, have important consequences
on our ability to detect quasars at $z\ga 7$, and is the primary focus of our work.
\begin{figure}
\centering
\includegraphics[scale=0.6]{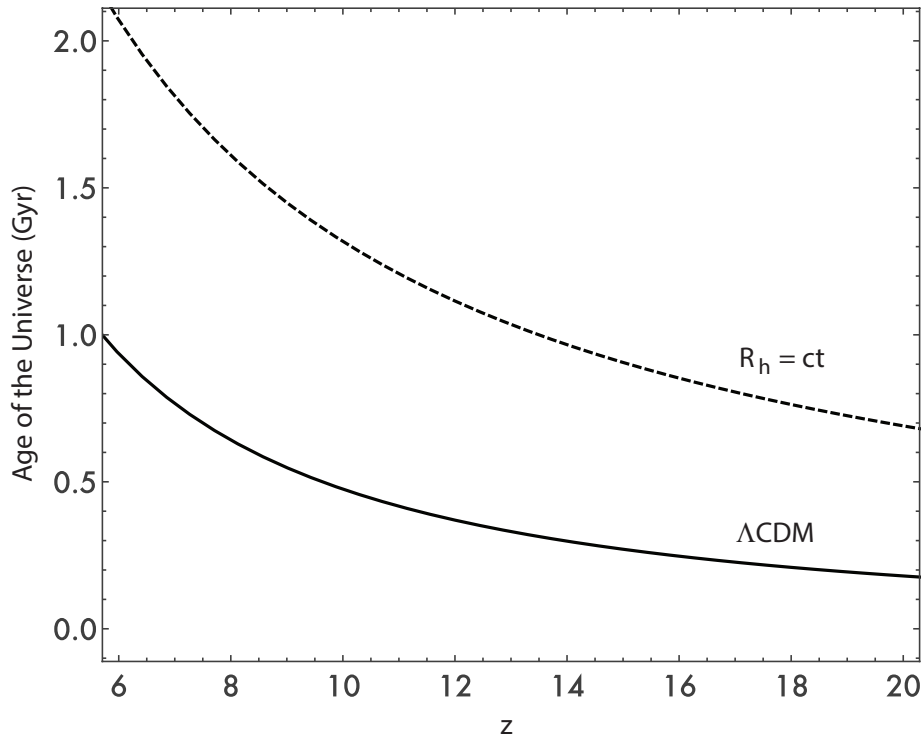}
\label{Fig 1}\caption{Age of the Universe versus redshift in both {\it Planck} $\Lambda$CDM (solid curve)
and $R_{\rm h} = ct$, assuming the same Hubble constant for simplicity (dashed curve).}
\end{figure}
\begin{figure}
\centering
\includegraphics[scale=0.6]{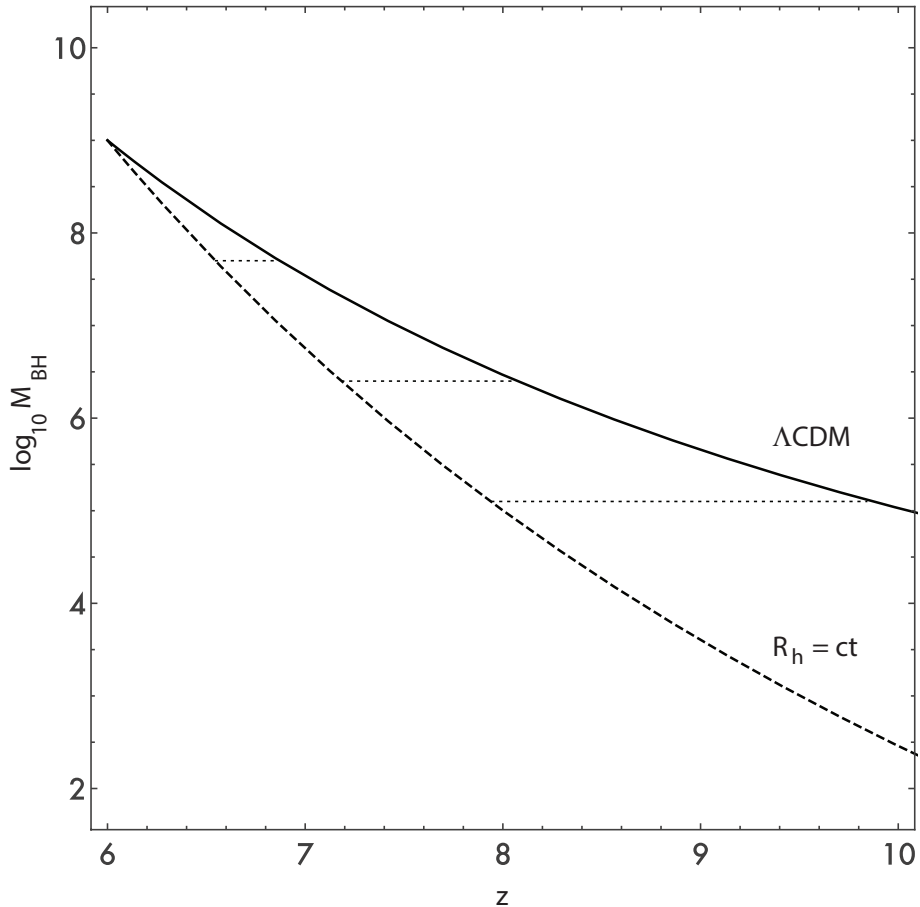}
\label{Fig 2}\caption{The mass evolution of a black hole with an observed mass
$M_0 =10^9 \;M_\odot$ at redshift $z_0 = 6$ in both $\Lambda$CDM  (solid curve) and
$R_{\rm h} = ct$ (dashed curve), for the benchmark values $\eps_r = 0.1$
and $\lambda_{\rm Edd} = 1$. The thin dashed lines denote where the Universe was $1/3\, t_{\rm Edd}$,
$2/3\, t_{\rm Edd}$ and $1\, t_{\rm Edd}$ younger than its age at $z = 6$ which, for the adopted
values of $\eps_r = 0.1$ and $\lambda_{\rm Edd} = 1$, correspond to a 3, 6 and 9 e-folding drop in mass.}
\end{figure}

\section{Emission Model}
\subsection{Disk Emission}
Our emission model follows the basic development in Pezzulli et al. (2017).
In the classical model, a geometrically thin disk has an inner radius given by the last stable orbit at radius
\be
r_0 = 3 R_{\rm S} = {6 \,G M_{\rm bh} \over c^2}\,,
\ee
and has the temperature profile
\be
T(r) = \left({3 G M_{\rm bh} \dot M\over 8 \pi\sigma r^3}\right)^{1/4} \,\left(1-\sqrt{{r_0\over r}}\right)^{1/4}\,,
\ee
for which the maximum temperature is achieved at $r = {49\over 36} r_0$
(Shakura \& Sunyaev 1973).  If the
disk emits a perfect blackbody, the bolometric luminosity is then given by
the well know expression
\be
L_{bol} = {1\over 12} \dot M c^2 \,,
\ee
which sets the radiative efficiency at $\eps_r = 1/12$.  One thus need only specify
the black hole mass and the bolometric
luminosity  (or alternatively, the mass accretion rate $\dot M$) in order to calculate the emission from the disk.

To allow for a more general treatment where both $L_{bol}$ and $\eps_r$
can be used as model parameters, we calculate the disk emission through the  expression
\be
L_\nu = L_0 \int_{r_i}^\infty B_\nu (T[r]) \,r\; dr\,,
\ee
where $B_\nu$ is the Planck function  and $L_0$ is a normalizing factor used to set
the bolometric luminosity
\be
L_{bol} = \int_0^\infty L_\nu\; d\nu\,.
\ee
The inner disk radius $r_i$ is determined by considering whether a thin disk or a slim disk serves as a better representation of the system under consideration (Abramowicz et al. 1988).  Specifically, for a disk to remain
geometrically thin, $L_{bol} \la 0.3\, L_{\rm Edd}$.  If this condition is not met, radiation pressure inflates
the disk, which is then better described by a slim disk model.  In this case, photons are trapped
for radii $r \la r_{pt} = 1.5 H (R_{\rm S} / r) (\dot M / \dot M_{\rm Edd})$, where $H$ is the
half-disk thickness.   Assuming $H / r = 2/3$, and since  $\dot M / \dot M_{\rm Edd} =
L_{bol} / L_{\rm Edd}$, we therefore set
$r_i = \hbox{Max}[3 R_{\rm S}, \lambda_{\rm Edd} R_{\rm S}]$.

\subsection{Soft X-ray emission}
X-ray surveys have proven to be suitable for identifying quasars at high redshift,
and are advantaged over lower wavelengths due to a smaller amount of obscuration
and less contamination or dilution from the host galaxy.
The X-ray emission originates from a hot corona that surrounds the disk (see also
Liu \& Melia 2001), and can be parametrized as a power-law with
exponential cutoff at $E_c = 300$ keV:
\be
L_{X,\nu} \propto \nu^{-\Gamma+1}  e^{-h\nu/E_c}\,,
\ee
where the photon index is set through the empirical relation
\be
\Gamma = 0.32 \log_{10}\left({L_{bol}\over L_{\rm Edd}}\right) + 2.27\,,
\ee
(Brightman et al. 2013).
We use the results of Lusso \& Risaliti (2106; fig.~6) to then normalize
the X-ray luminosity via the expression
\be
\log_{10} L_{2\;{\rm keV}} = 0.638 \log_{10} L_{2500} + 7.074\,,
\ee
where both luminosity densities are in units of erg s$^{-1}$ Hz$^{-1}$.  The disk
(solid) and X-ray (dashed) emission for systems with $M_{\rm bh} = 10^6\;M_\odot$
and $10^9\;M_\odot$, both with $\lambda_{\rm Edd} = 1$ and with $\eps_r = 0.1$, are shown in figure~3.

\begin{figure}
\centering
\includegraphics[scale=0.6]{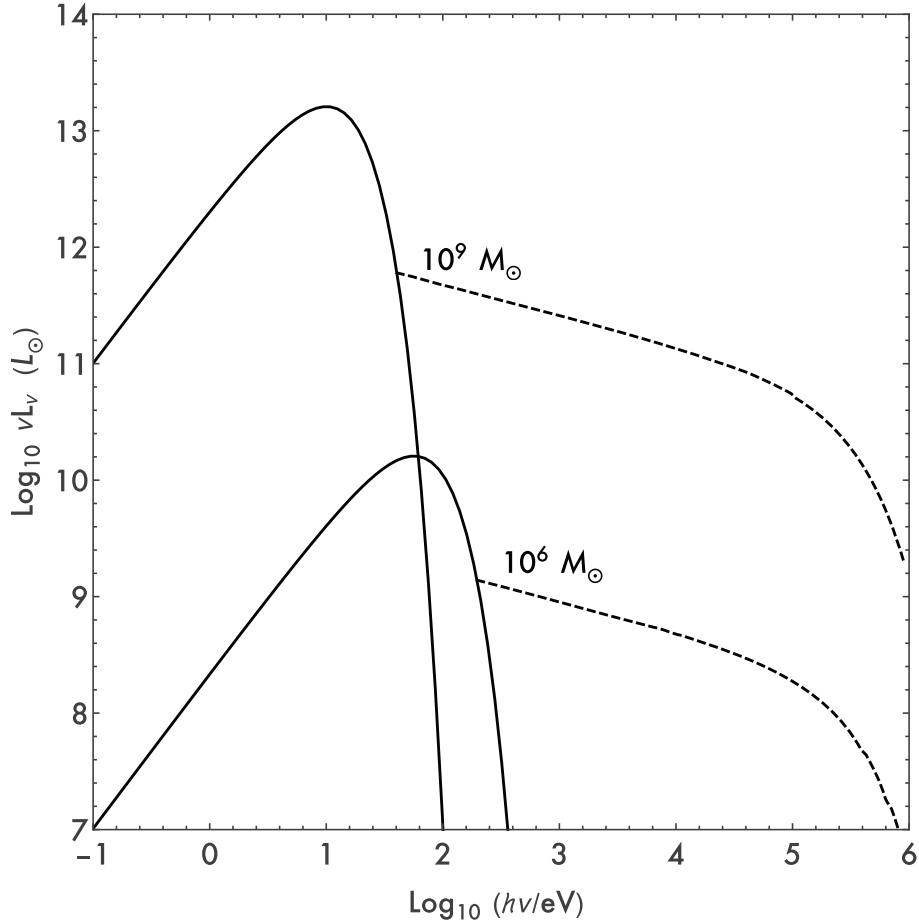}
\label{Fig 3}\caption{Emission from the disk (solid curve) and hot corona (dashed) of black holes with masses
of $10^6$ and $10^9\;M_\odot$.}
\end{figure}

Since the greatest sensitivity to the emission produced in our model occurs in soft X-rays
(see, e.g., Pezzulli et al. 2017), we consider only the soft X-ray band between $\eps_l = 0.5$ keV and
$\eps_u = 2$ keV. For a black hole at redshift $z$, the luminosity emitted in this band is
given by the expression
\be
L_{X} = \int_{\eps_l (1+z)}^{\eps_u (1+z)} L_{X,\nu}\; d\nu\,.
\ee
The flux received at Earth is then given by
\be
F_X = {L_X \over 4\pi D_L^2} \,,
\ee
where $D_L$ is the luminosity distance, which in $\Lambda$CDM and $R_{\rm h}=ct$ are given, respectively,
by the expressions
\be
D^{\Lambda{\rm CDM}} = {c\over H_0 } (1+z)\int_0^z {du\over\sqrt{\Omega_m (1+u)^3 +
\Omega_r (1+u)^4 + \Omega_\Lambda (1+u)^{3+3w}}}\;,
\ee
and
\be
D^{R_{\rm h}=ct} = {c\over H_0}   (1+z)  \ln (1+z)\;.
\ee
We note that these distances are within $10\%$ of each other throughout the
$z =10$ to $z = 6$ epoch (see fig.~4), indicating that the observational differences
of quasars during this epoch are due almost entirely from the difference in the temporal
evolution (and subsequently, the mass growth) between the two cosmologies. In addition,
since the luminosity distances are very similar at $z \sim 6$, we do not compensate
for observationally determined values of luminosity derived assuming $\Lambda$CDM
when using those results in $R_{\rm h} = ct$.
\begin{figure}
\centering
\includegraphics[scale=0.6]{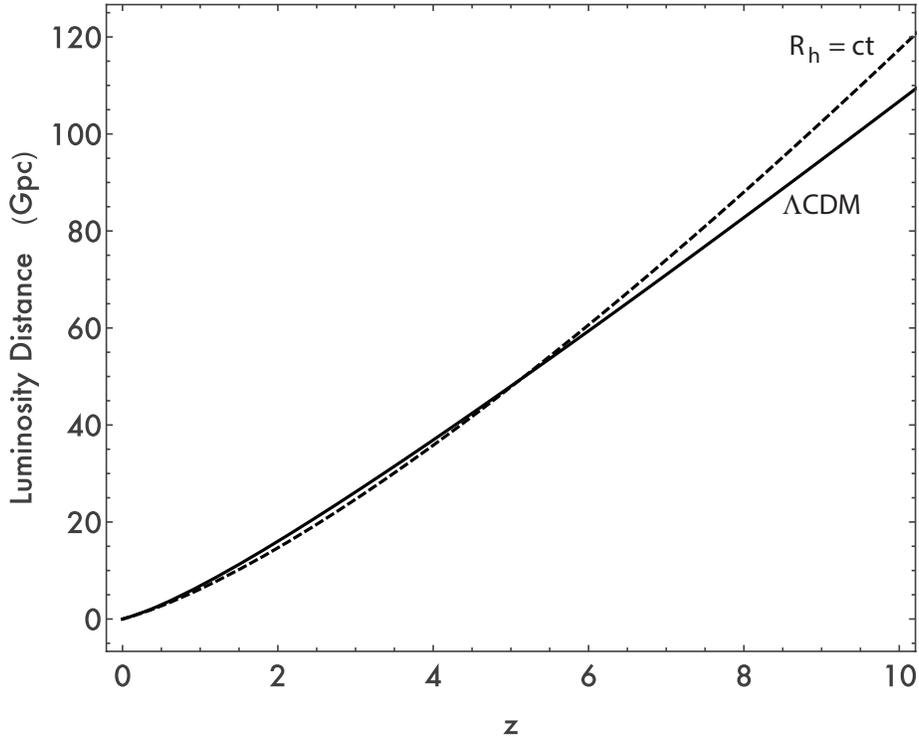}
\label{Fig 4}\caption{The luminosity distance as a function of redshift for $\Lambda$CDM (solid curve) and
$R_{\rm h} = ct$ (dashed curve).}
\end{figure}

We ignore the Compton reflection of X-rays by the disk since the effect is minimal in the
soft X-ray band (Magdziarz \& Zdziarski 1995; Markoff, Melia \& Sarcevic 1997; 
Trap et al. 2011).  To keep the analysis as simple as possible,
we also ignore absorption due to interactions with the surrounding gas and dust.   Our flux
calculations and subsequent estimates on the number of observable quasars should therefore be
taken as upper limits.  However, the results of Pezzulli et al. (2017) indicate that even
if absorption is important, the overall effect out to $z \sim 10$ is not expected to
be severe enough to impact our results.

\section{Black-hole evolution and the resulting X-ray flux}
We now combine the mass-growth model developed in \S2 with the emission model
from \S 3 in order to determine the observable flux of black holes evolving in $\Lambda$CDM and
$R_{\rm h} = ct$. For illustrative purposes, we first apply our model to the sample of observed
quasars highlighted in Nanni et al. (2017) with the best counting statistics in X-rays:  
J0100+2802, J1030+0524, J1120+0641, J1148+5251, and J1306+0356.  The values of
$z$, $M_{BH}$, and $\lambda_{\rm Edd}$ for these sources are reproduced in 
Table~1. The adopted mass for J1120+0641 is based on observations of the
MgII line, while the mass and Eddington ratio for J1030+0524 and J1306+0356
were obtained by averaging the results presented in Table~4 of de Rosa et al. (2011),
based on the use of their Equation~4. The radiative efficiency is 
assumed to be $\eps_r = 0.1$ in all cases.  We calculate the observed flux in the 
$0.5 - 2$ keV soft X-ray band as a function of redshift. The color coded results are 
shown in figure~5 for $\Lambda$CDM and figure~6 for $R_{\rm h}= ct$, with data points 
representing  the $0.5 - 2$ keV fluxes derived by Nanni et al. (2017) based on 
{\it Chandra} observations.

\begin{figure}
\centering
\includegraphics[scale=0.7]{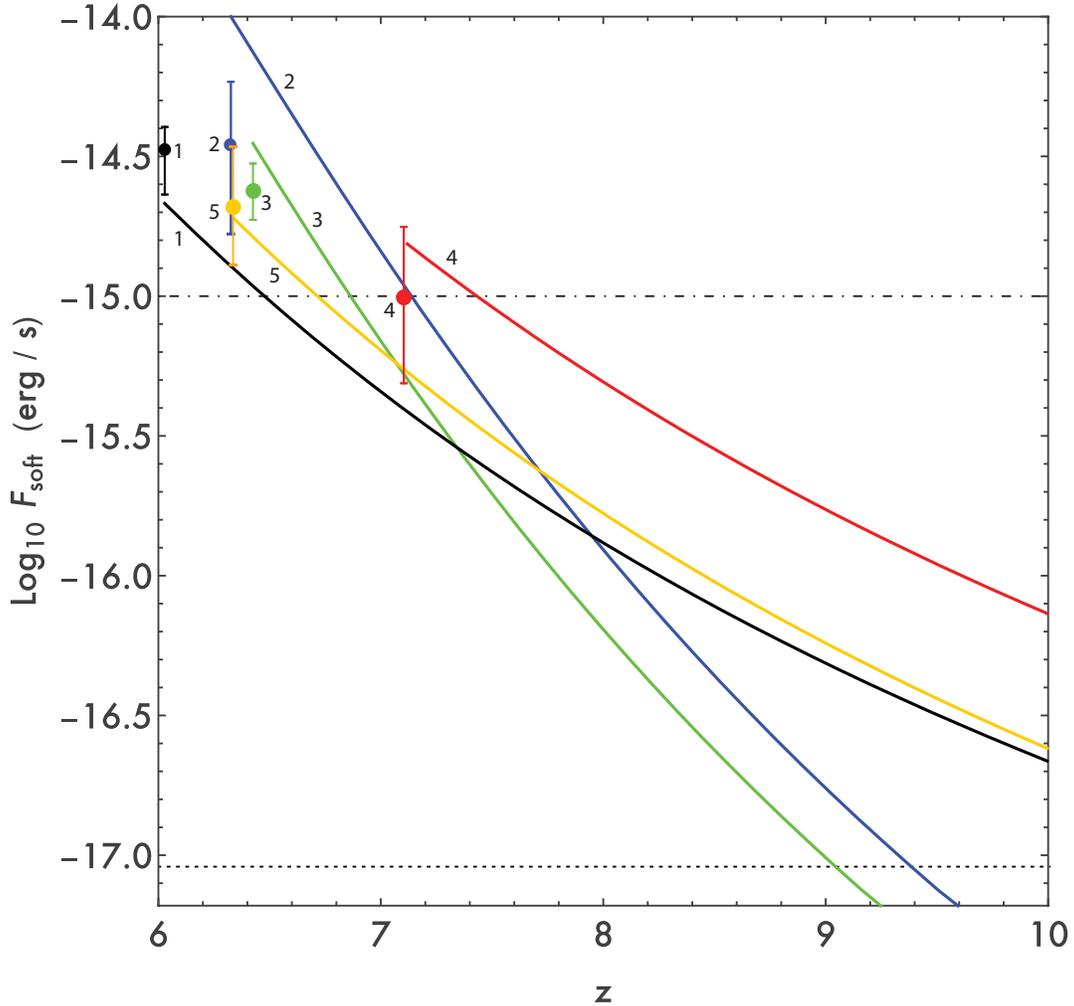}
\label{Fig 5}\caption{Soft X-ray band ($0.5 - 2$ keV) flux observed at Earth as a function of redshift
in our $\Lambda$CDM evolutionary scenario, for 5 observed quasars highlighted in Nanni et al. (2017) 
with the best counting statistics in X-rays:  J0100+2802 (blue,2), J1030+0524 (orange,5), 
J1120+0641 (red,4), J1148+5251 (green,3), and J1306+0356 (black,1). Data points represent 
the  $0.5 - 2$ keV fluxes derived by Nanni et al. (2017) based on {\it Chandra} observations, 
together with their $1\sigma$ error bars. The dotted line represents the Chandra Deep Field South 
soft X-ray band flux limit of $9.1 \times 10^{-18}$ erg/s, while the dot-dashed line represents 
the lowest observed soft X-ray flux in our sample of quasars taken from Nanni et al. (2017).}
\end{figure}

\begin{table}
\caption{X-ray Detected Quasars}
\centering
\begin{tabular} {c c l l l}
\hline
\hline
Quasar & $z$ & $M_{BH}$ ($M_\odot$)& $\lambda_{Edd}$ & Reference\\ [0.5ex]
\hline
J1306+0356 & 6.0 & $2.3\times 10^9$ & 0.45 & De Rosa et al. (2011)\\
J0100+2802 & 6.3 & $1.2\times 10^{10}$ &  1.06 & Wu et al. (2015)\\
J1030+0524 & 6.3 & $2.2\times 10^9$ &0.5 & De Rosa et al. (2011) \\
J1148+5251 & 6.4 & $3.0\times 10^9$ &1.0 & Willott et al. (2003)\\
J1120+0641 & 7.1 & $2.4\times 10^9$ &0.5  & De Rosa et al. (2014)\\
\hline
\end{tabular}
\label{table:quasar}
\end{table}

\begin{figure}
\centering
\includegraphics[scale=0.7]{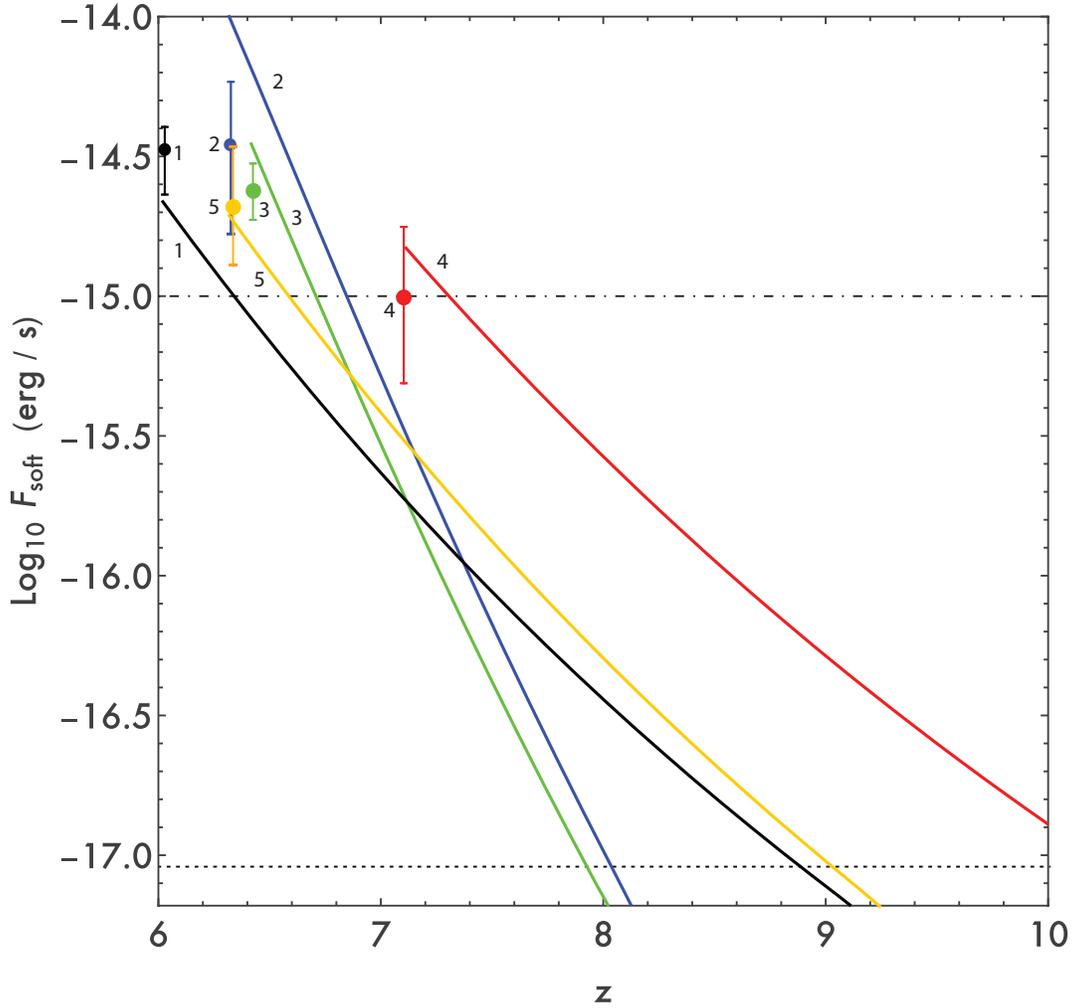}
\label{Fig 6}\caption{Soft X-ray band ($0.5 - 2$ keV) flux observed at Earth as a function of redshift
in $R_{\rm h} = ct$, for 5 observed quasars highlighted in Nanni et al. (2017) with the best counting 
statistics in X-rays: J0100+2802 (blue,2), J1030+0524 (orange,5), J1120+0641 (red,4), J1148+5251 (green,3), 
and J1306+0356 (black,1). Data points represent the $0.5 - 2$ keV fluxes derived by Nanni et al. (2017) 
based on $Chandra$ observations, together with their $1\sigma$ error bars. The dotted line 
represents the Chandra Deep Field South soft X-ray band flux limit of $9.1 \times 10^{-18}$ erg/s, 
while the dot-dashed line represents the lowest observed soft X-ray flux in our sample of quasars 
taken from Nanni et al. (2017).}
\end{figure}

Our results, which are in fairly good agreement with the observations, 
indicate that the difference in timelines between $\Lambda$CDM and
$R_{\rm h}= ct$ has clear implications for the observability of quasars between redshifts $6 - 10$.
In $\Lambda$CDM, assuming the quasars used in our analysis are representative of the broader
population, a significant fraction of their counterparts would produce a soft
X-ray flux above the lowest observed flux in our sample (dot-dashed line) between
redshifts $z \approx 7 -  7.5$, and would produce a soft X-ray flux above the Chandra
Deep Field South limit (dotted line) out to $z\sim 10$. The situation is considerably
different in $R_{\rm h} = ct$, where relatively few quasars would be detected above our
established sample threshold (dot-dashed line) out to a redshift of $z \approx7.5$,
and detection at the Chandra Deep Field South limit would be rare for $z \ga 10$.

While figures~5 and 6 provide an important insight into how SMBHs evolve in both cosmologies,
we wish to put our analysis on a firmer statistical footing. The recent (and ongoing) discovery,
imaging and spectroscopic analysis of quasars at $z \approx 6$ now allows us to carry out
a statistical analysis of their properties over a range of luminosities, and has lead to
a determination of the quasar mass function at $z = 6$. Specifically, the analysis
of Willott et al. (2010), based on the absolute magnitude at 1450 $\AA$ for a sample of
$z\approx6$ quasars, yielded a Schechter mass function of the form
\be
\Phi_M[M_{\rm bh}; z=6] = \Phi_0 \left( {M_{\rm bh} \over M^*}\right)^\alpha e^{-M_{\rm bh}/M^*}\,,
\ee
with best fit parameters $\Phi_0 = 1.23\times 10^{-8}$ Mpc$^{-3}$ dex$^{-1}$,
$M^* = 2.24\times 10^9$ $M_\odot$, and $\alpha = -1.03$.
This mass function normalized to yield the number of quasars
$dN(M_{\rm bh},z)$ per mass dex between redshift $z$ and $z+dz$, evaluated at $z = 6$, is shown in figure~7
for both cosmologies.

\begin{figure}
\centering
\includegraphics[scale=0.6]{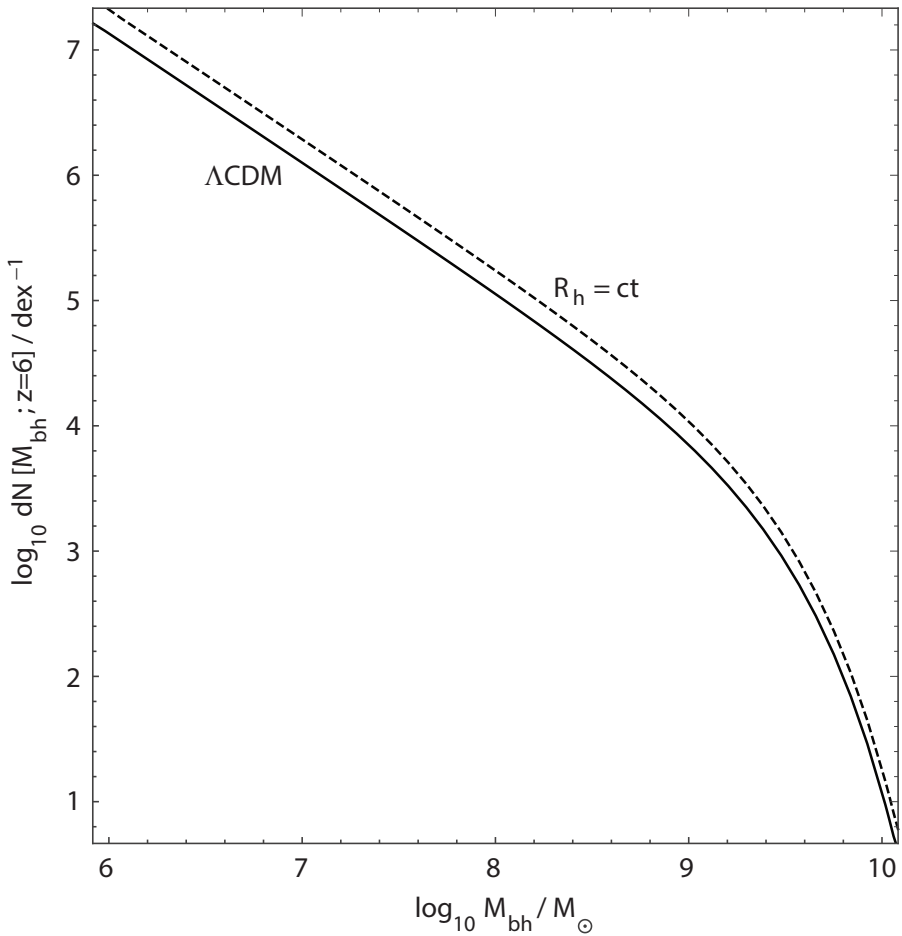}
\label{Fig 7}\caption{The mass function normalized to yield the number of quasars
$dN(M_{\rm bh},z)$ per mass dex between redshift $z$ and $z+dz$, evaluated at $z = 6$,
for both cosmologies.}
\end{figure}

To keep our analysis as direct as possible, we use the fairly narrow distribution in
$\lambda_{\rm Edd}$ values observed at $z \approx6$ (see fig.~6 in Willott et al. 2010)
as justification for setting $\lambda_{\rm Edd} = 1$ for
all quasars in the early Universe.  In the absence of mergers, and with accretion occurring
at the Eddington rate, all black holes evolve lock-step during the early Universe, and the
mass function at any redshift can be easily obtained through the transformation
\be
\Phi_M[M_{\rm bh}; z] = \Phi_M[M_{\rm bh} e^{{(1-\eps_r)\over\eps_r} { \lambda_{\rm Edd} \over t_{\rm Edd}} (t_6 - t_z)}\,;z=6] \,,
\ee
where $t_z$ is the age of the Universe at redshift $z$ (see Eqns.~ 2--4).

In order to assess the detectability of quasars at higher redshift for both cosmologies
under consideration, we convert the luminosity function to a flux function
at values of $z$ = 6, 7 and 8 for each model.  The results are shown in figure~8, with
the solid curves representing the $\Lambda$CDM case and the dashed curves
representing the $R_{\rm h} = ct$ case. Note that the slight mismatch at $z$ = 6
results from the slight difference in luminosity distance between the two cases,
which, as discussed above, was not corrected for. As can be clearly seen, a
significantly smaller fraction of the quasars detected at $z = 6$ can also be seen
at higher redshifts for $R_{\rm h} = ct$ than for $\Lambda$CDM. This result stems
almost entirely from the difference in mass growth between the two scenarios. For
example, a time of 0.17 Gyr passes between $z = 7$ and $z = 6$ in $\Lambda$CDM,
while the corresponding span of time is 0.26 Gyr in $R_{\rm h}= ct$. At $z = 6$,
a SMBH with mass $M_{\rm bh} = 1.5\times 10^6\;M_\odot$ produces a soft X-ray flux at the
Chandra limit.  That result does not change much at $z = 7$ for either Universe, since the
luminosity distance changes by less than 25\%.  However, a SMBH evolving from $z = 7$ to
$z = 6$ increases its mass by a factor of 30 in $\Lambda$CDM, and a factor of 180 in
$R_{\rm h} = ct$. The corresponding shift in the mass function, as expressed by
Equation~(18), is therefore much greater in the $R_{\rm h} = ct$ universe.

\begin{figure}
\centering
\includegraphics[scale=0.7]{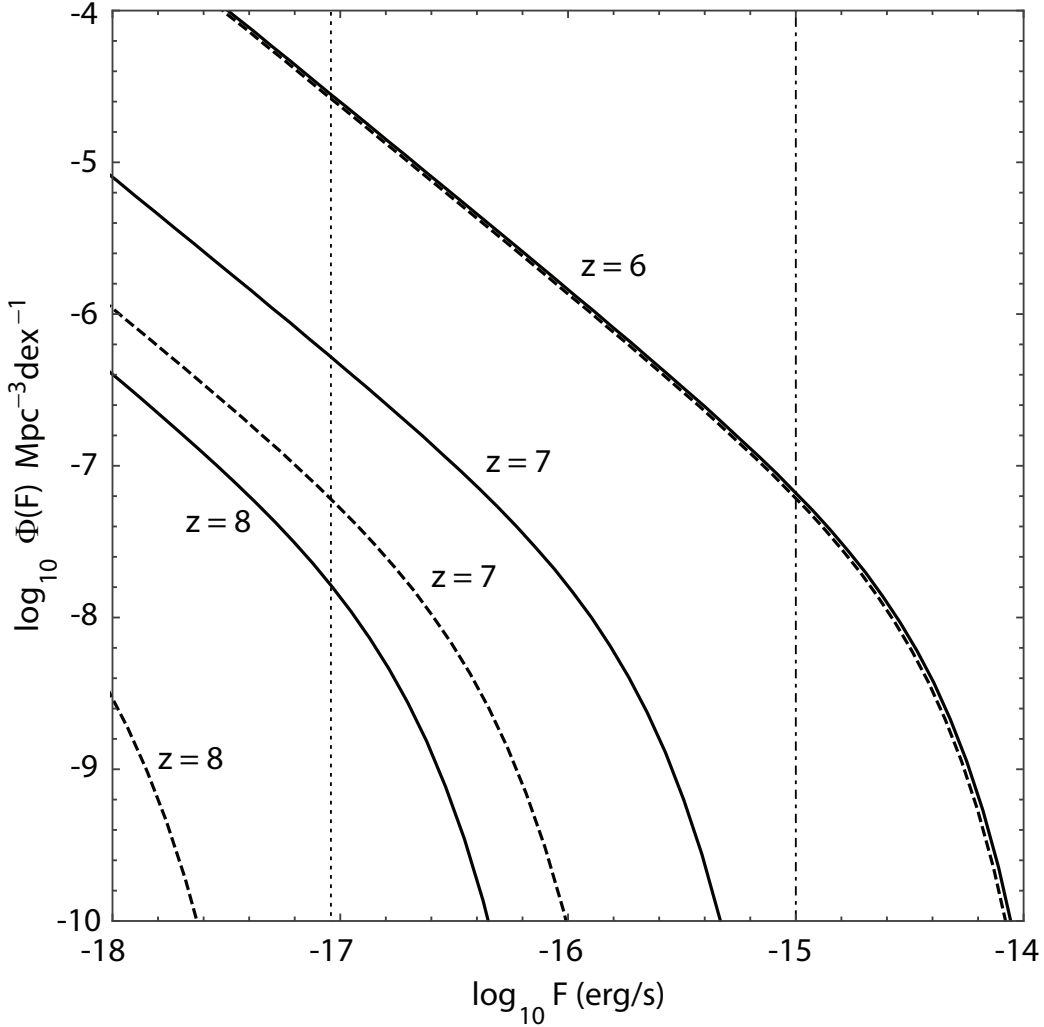}
\label{Fig 8}\caption{Flux distributions for $\Lambda$CDM (solid) and $R_{\rm h} = ct$ (dashed)
at $z = 6, 7$ and 8. The dotted line represents the Chandra Deep Field South soft X-ray band
flux limit of $9.1 \times 10^{-18}$ erg/s, and the dot-dashed line represents the observed
soft X-ray flux of the lowest mass SMBH in the sample used to generate figures~5 and 6, as
calculated by our model.}
\end{figure}

To further illustrate this point, we next determine what fraction of quasars
that produce a flux above the Chandra soft X-ray band limit at $z = 6$ would still do so
at higher redshifts. As noted above, a black hole at $z = 6$ would need a mass of
$M_{\rm bh} = 1.5 \times 10^6\;M_\odot$ to produce a flux equal to the Chandra
Deep Field South soft X-ray band flux limit in our accretion model. Taking as our
parent population all black holes at $z = 6$ with mass greater than this limit,
we then find the corresponding masses for which the population of black holes with a greater
mass represent 80, 60, 40, 20, 10, 1 and 0.1\%, respectively, of the parent population. 
For each of these
demarking masses, we then calculate the flux evolution, as was done for figures~5 and 6,
using the same parameters $\eps_r = 0.1$ and $\lambda_{\rm Edd} = 1$.  The results
are displayed in figure~9 for $\Lambda$CDM and figure~10 for $R_{\rm h} = ct$.
In $\Lambda$CDM, 10\% of our parent population would be observable (above the
Chandra limit) beyond a redshift of $z\approx 6.55$, and only
around 1\% would be observable beyond redshift $z \ga 7$.  In contrast, for $R_{\rm h} = ct$,
10\% of our parent population would be observable beyond a redshift of $z
\approx 6.35$, and the most massive 1\% would be observable beyond a
redshift of $z \approx 6.7$. These results indicate that the next generation of
observations, which should be able to provide a statistically significant number
of $z \ga 7$ detections, will be able to differentiate between $\Lambda$CDM
and $R_{\rm h} = ct$. In both cases, only the most massive 0.1\% of quasars 
produce a flux comparable to the minimum observed flux from our Nanni et al. (2017) 
sample used for Figures 5 and 6 (dot-dashed line) at  a redshift of $z \sim 6$.

\begin{figure}
\centering
\includegraphics[scale=0.7]{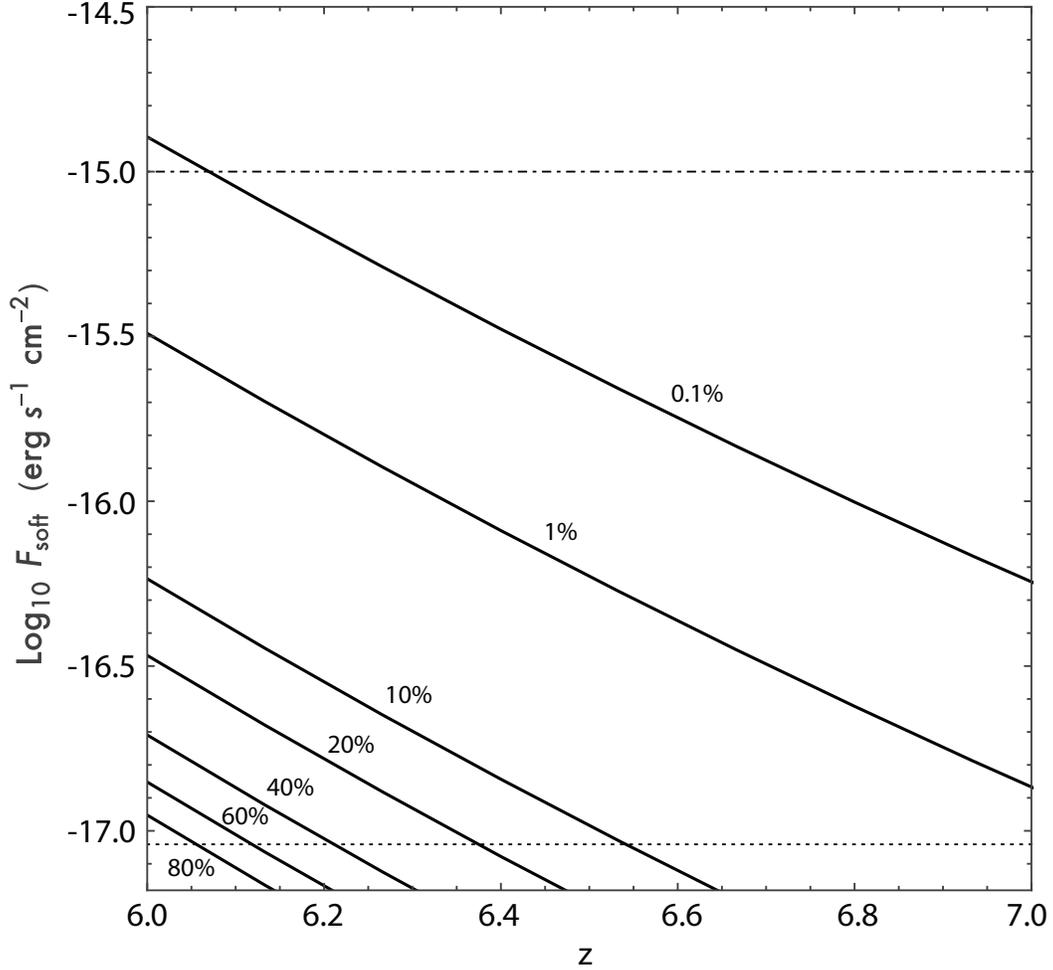}
\label{Fig 9}\caption{Chandra soft X-ray flux versus $z$ tracks in $\Lambda$CDM for quasars with black hole
masses that demark 80, 60, 40, 20, 10, 1 and 0.1\% of the parent population of all quasars with a mass
greater than $1.5 \times 10^6\;M_\odot$ at $z = 6$, corresponding to the mass limit that produces
a flux at that redshift equal to the Chandra Deep Field South soft X-ray band flux limit of
$9.1 \times 10^{-18}$ erg/s (indicated by the dotted line). The dot-dashed line represents
the observed soft X-ray flux of the lowest mass SMBH in our sample used for figs.~5 and 6.}
\end{figure}

\begin{figure}
\centering
\includegraphics[scale=0.7]{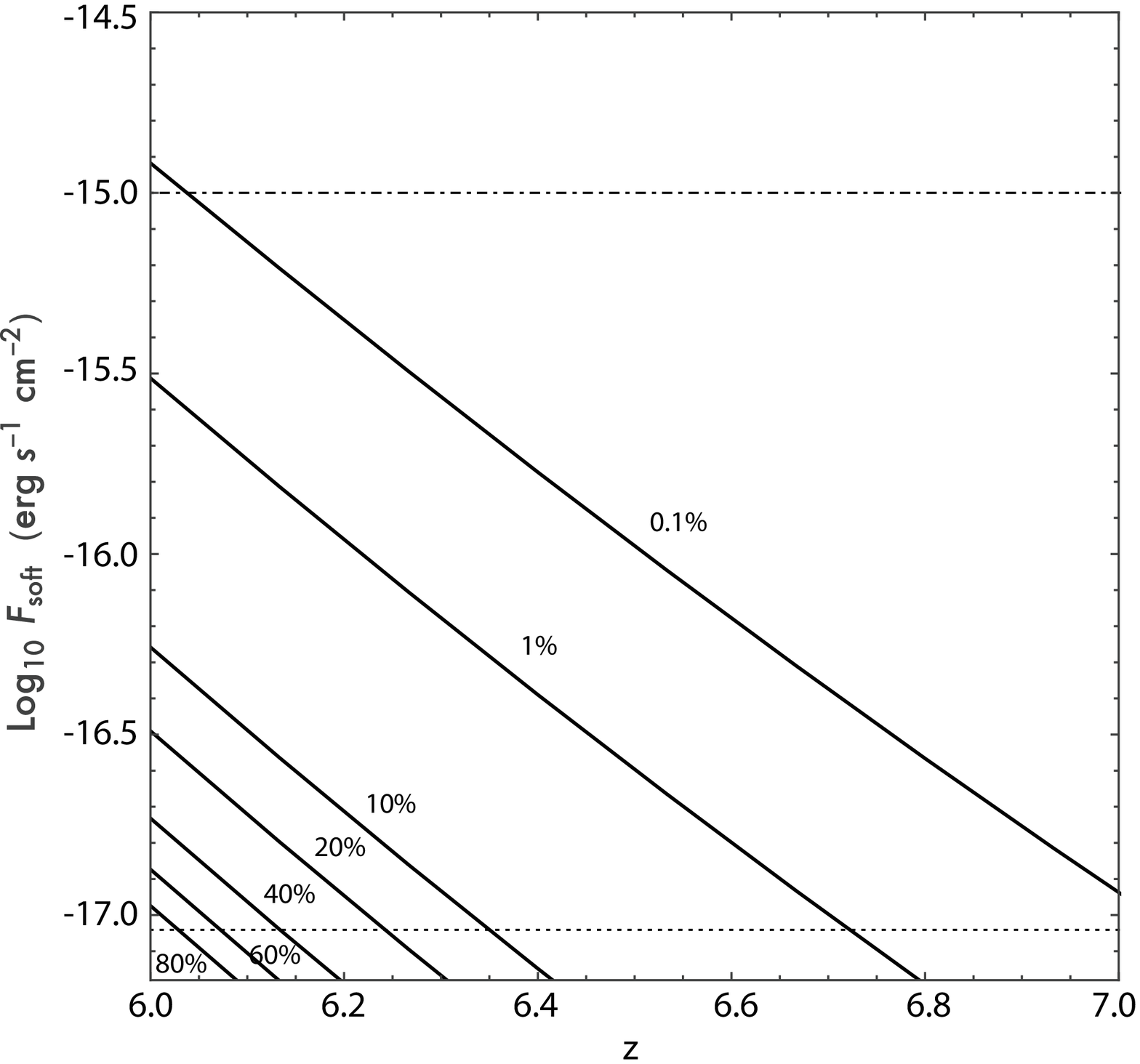}
\label{Fig 10}\caption{Chandra soft X-ray flux versus $z$ tracks in $R_{\rm h} = ct$ for quasars with black hole
masses that demark 80, 60, 40, 20, 10, 1 and 0.1\% of the parent population of all quasars with a mass greater
than $1.5 \times 10^6\;M_\odot$ at $z = 6$, which corresponds to the mass limit that produces
a flux at that redshift equal to the Chandra Deep Field South soft X-ray band
flux limit of $9.1 \times 10^{-18}$ erg/s  (indicated by the dotted line).
The dot-dashed line represents the observed soft X-ray flux of the lowest
mass SMBH in our sample used for figures~5 and 6.}
\end{figure}

We conclude this analysis by estimating how many quasars should be detectable above a given
flux threshold as a function of redshift by integrating the quasar mass function above that
threshold out to $z = 10$. This number is given by the integral expression
\be
N(\ge F; z) \equiv \int_{z}^{10} \int_{\log_{10}[M_t(z')]}^\infty \Phi(M_{\rm bh},z') V_{z'}\, d [\log_{10}(M_{\rm bh})]\,  dz'\,,
\ee
where $V_z = 4\pi D_c^2 \,dD_c/dz$ is the comoving differential volume and $M_t(z)$ represents
the black hole mass at redshift $z$ required to produce a soft X-ray flux equal to $F$.
Note that the comoving distance $D_c$ is smaller than the luminosity distance by a factor
$(1+z)$. The results are presented in figure~11. As noted already, there is a significant
difference in the expected number of detectable quasars predicted by the two cosmological
models.
\begin{figure}
\centering
\includegraphics[scale=0.7]{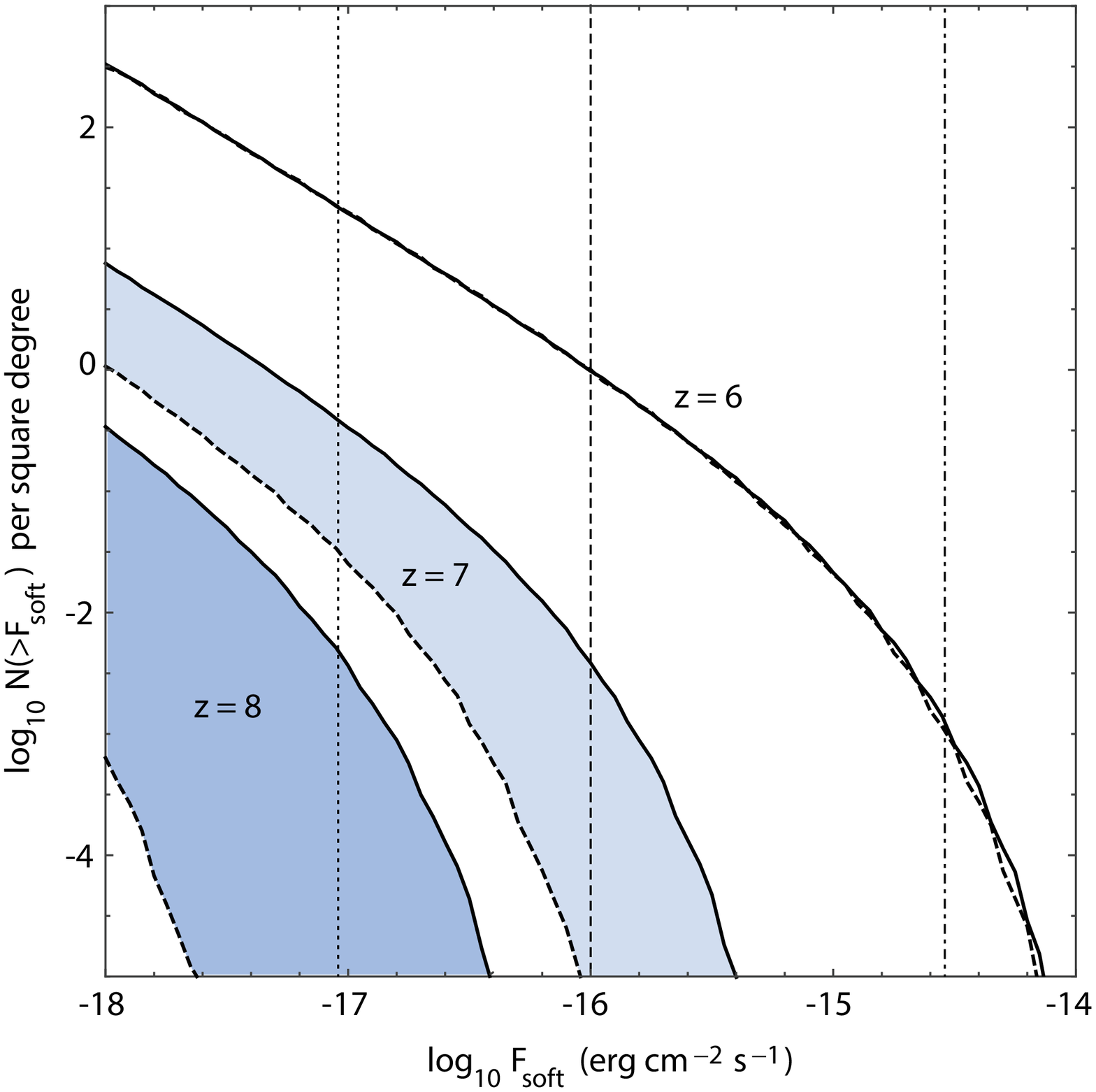}
\label{Fig 11}\caption{Number (density per square degree) of quasars that should be detectable
above a given flux threshold $F_{\rm soft}$ as a function of that threshold for
$z = 6$, 7, and 8. At $z=6$, both $\Lambda$CDM and $R_{\rm h}=ct$ predict the same
value because the curves are normalized to the observed luminosity function at that redshift.
At larger redshifts, $\Lambda$CDM is indicated by the solid curve, while $R_{\rm h}=ct$
is shown with a dashed curve. The colored regions emphasize the differences between
these two models at $z=7$ and 8. The dotted line represents the Chandra Deep Field South
soft X-ray band flux limit of $9.1 \times 10^{-18}$ erg/s, the dashed line
represents the expected Athena flux limit, and the dot-dashed line represents the soft
band (0.5 - 2 keV)  point-source sensitivity for eROSITA at $\sim 140$ deg$^2$.}
\end{figure}

Based on the observed luminosity function at $z=6$, which is used to normalize the
expected number of detectable quasars in both models, our analysis indicates that
there is a steep drop off in the number of quasars that can be observed at $z\sim 7$
compared to $z\sim 6$ at flux limits greater than $\ga 10^{-15}$ erg s$^{-1}$ cm$^{-2}$.
As noted above, a SMBH evolving from $z = 7$ to $z = 6$ increases its mass by a factor 
of 30 in $\Lambda$CDM, and a factor of 180 in $R_{\rm h} = ct$.  Similar to what was 
seen in Figure 8, the result is a more pronounced downward (leftward) shift (by about 
an order of magnitude) in the $R_h = ct$ curve than its $\Lambda$CDM counterpart at $z = 7$
compared to their common $z = 6$ curves.   But the turnover in the mass distribution function
acts to amplify the observational consequences between the two cosmologies. With a flux 
sensitivity $\ga 3 \times 10^{-15}$ erg s$^{-1}$ cm$^{-2}$ (represented by the dot-dashed
vertical line in Figure 11), it seems very unlikely that eROSITA will be able to provide 
the observational evidence needed to discriminate between the two cosmological models under
consideration.  In contrast, the roughly three orders of magnitude difference in the 
number of $z\sim 7$ quasars that can be observed in $R_{\rm h}=ct$ versus $\Lambda$CDM 
for the ATHENA flux sensitivity (represented by the dashed vertical line
in Figure 11) makes it quite likely that the observations made by that instrument will be able to
discriminate between the two cosmologies in the next decade. Specifically, while
we expect that ATHENA will detect very few, if any, quasars at $z \sim 7$ in $R_h = ct$,
it should detect several hundred of them in $\Lambda$CDM---a rather compelling quantitative
difference.
\newpage
\section{Conclusions}
The detection of billion-solar-mass quasars at $z\ga 6$ has created some
tension with the {\it Planck} $\Lambda$CDM model, in the sense that conventional
Eddington-limited accretion, as we understand it in the local Universe, could not
have produced such large objects in the scant $400-500$ Myr afforded them by the
timeline in this cosmology. Remedies to circumvent this problem have included
models to create $\sim 10^5\;M_\odot$ seeds or to permit transient super-Eddington
accretion, requiring a very low duty-cycle. Both of these solutions are anomalous
because no evidence for either of them has ever been seen. In fact, the quasar
luminosity function towards $z\sim 6$ suggests that the inferred accretion rate
saturates at close to the Eddington value, with a spread no greater than about
0.3 dex. The data seem to be telling us that these supermassive black holes
probably grew to their observed size near $z=6$ by accreting more or less steadily
at roughly the Eddington value.

In previous work, we demonstrated that an alternative, perhaps more elegant, solution
to this problem may simply be to replace the timeline in $\Lambda$CDM with that
in the $R_{\rm h}=ct$ universe. By now, these two models have been compared with
each other and tested against the observations using over 20 different kinds of data.
The $R_{\rm h}=ct$ model has not only passed all of these tests, but has actually
been shown to account for the data analyzed thus far better than the standard model.
There is therefore ample motivation to advance the study of SMBH evolution in this
cosmology beyond mere demographics.

This has been the goal of this paper---to examine the progenitor statistics in
both models, based on the observed luminosity function at $z\sim 6$. We have
sought to keep the analysis as simple and straightforward as possible, avoiding
unnecessarily complicated SED contributions. For this purpose, the $0.5-2.0$ keV
flux, thought to be produced in the corona overlying the accretion disk, appears
to be an ideal spectral component. The model for producing this emissivity is
simple, and probably reliable over a large range in black-hole mass. In addition,
one can easily compensate for a transition in the accretion rate, from low to
large values.

With this basic accretion model, we have demonstrated that---for the two
cosmologies examined here---the difference in the expected number of detections 
with the upcoming ATHENA mission is very large. According to figure~11, and based on 
the observed luminosity function at $z\sim 6$, the ATHENA mission is expected
to detect approximately $3.9\times 10^{-6}$ quasars per square degree at 
$z\sim 7$ in $R_{\rm h}=ct$, i.e., approximately $0.16$ over the whole sky.
By comparison, this number is $\sim 3.9\times 10^{-3}$ quasars per square
degree at $z\sim 7$ in $\Lambda$CDM, or roughly $160$ over the whole sky.

The caveat, of course, is that we have ignored the impact of mergers throughout
this analysis. Superficially, one could reasonably expect that the merger rate
should be about the same in both cosmologies. Nonetheless, the numbers shown
in figure~11 should be viewed as upper limits. One also needs to take into
account the fact that an observed detection rate lower than that predicted by
$\Lambda$CDM may be partially due to SMBH growth via these mergers, rather than
it being a strong indication that the timeline in $R_{\rm h}=ct$ is preferred
by the data. On the flip side, mergers cannot lead to a detection rate much
higher than that shown in figure~11, depending on how reliable our streamlined
accretion model happens to be. Thus, if the number of high-$z$ quasars detected
with future surveys is closer to that predicted by $\Lambda$CDM (i.e., the solid
curves in this figure), this would argue strongly against $R_{\rm h}=ct$, particularly 
at $z\sim 8$, where the difference is expected to be even more pronounced.

\acknowledgments
We are grateful to the anonymous referee for several important suggestions
that have led to an improved presentation of our results. FM is also grateful 
to the Instituto de Astrof\'isica
de Canarias in Tenerife and to Purple Mountain Observatory in Nanjing, China
for their hospitality while part of this research was carried out. FM is also
grateful for partial support to the Chinese Academy of Sciences Visiting
Professorships for Senior International Scientists under grant 2012T1J0011,
and to the Chinese State Administration of Foreign Experts Affairs under
grant GDJ20120491013. MF is supported at Xavier University through the 
Hauck Foundation.


\begin{thebibliography}

\bibitem[1]{} Abramowicz M. A., Czerny B., Lasota J. P., Szuszkiewicz E., 1988, ApJ, 332, 646
\bibitem[2]{} Ade, P.A.R. et al., 2016, A\&A, 594, id A13
\bibitem[3]{} Brightman, M., et al., 2013, MNRAS, 433, 2485
\bibitem[4]{} Chan, C.-K., Liu, S., Fryer, C. L., Psaltis, D., \"Ozel, F., Rockefeller, G. \& Melia, F. 2009, ApJ, 701, 521
\bibitem[5]{} de Rosa, G., et al. 2011, ApJ, 739, 56
\bibitem[6]{} de Rosa, G., et al. 2014, ApJ, 790, 145
\bibitem[7]{} Fiore, F., et al., 2009, ApJ, 693, 447
\bibitem[8]{} Georgakakis, A., et al., 2015, MNRAS, 453, 1946
\bibitem[9]{} Liu, S. \& Melia, F., 2001, ApJL, 561, L77
\bibitem[10]{} Lusso, E., Risaliti, G., 2016, ApJ, 819, 154
\bibitem[11]{} Madau, P., Haardt, F., Dotti, M., 2014, ApJ, 784, L38
\bibitem[12]{} Magdziarz, P., Zdziarski, A. A., 1995, MNRAS, 273, 837
\bibitem[13]{} Markoff, S., Melia, F. \& Sarceivc, I., 1997, ApJL, 489, L47
\bibitem[14]{} Melia, F., 2013, ApJ, 764, 72
\bibitem[15]{} Melia, F., 2014, A\&A, 561, A80
\bibitem[16]{} Melia, F., 2017, MNRAS, 464, 1966
\bibitem[17]{} Melia, F. \& Fatuzzo, M., 2016, MNRAS, 456, 3422
\bibitem[18]{} Melia, F. and Maier, R. S., 2013, MNRAS, 432, 2669
\bibitem[19]{} Melia, F. and McClintock, T. M., 2015, Proc. R. Soc. A, 471, 20150449
\bibitem[20]{} Melia, F., Wei, J.-J., Wu, X.-F., 2015, AJ, 149, 2
\bibitem[21]{} Mortlock, D. J. et al., 2011, Nature, 474, 616
\bibitem[22]{} Nanni, R., Vignali, C, Gilli, R., Moretti, A. and Brandt, W. N., 2017, arXiv: 1704.08693
\bibitem[23]{} Pezzulli, E., Valiente, R., Orofino, M., Schneider, R., Sbarrato, T., 2017, MNRAS, 466, 2131
\bibitem[24]{} Ruffert, M. \& Melia, F., 1994, A\&A, 288L, L29
\bibitem[25]{} Shakura N. I., Sunyaev R. A., 1973, A\&A, 24, 337
\bibitem[26]{} Trap, G., Goldwurm, A., Dodds-Eden, K., Weiss, A., Terrier, R., Ponti, G., Gillessen, S., Genzel, R., Ferrando, P., B\'elanger, G. et al.
2011, A\&A, 528, id.A140
\bibitem[27]{} Treister, E., Schawinski, K., Volonteri, M., Natarajan, P., 2013, ApJ, 778, 130
\bibitem[28]{} Volonteri, M. and Rees, M. J., 2005, ApJ, 633, 624
\bibitem[29]{} Volonteri, M., Silk, J. and Dubus, G., 2015, ApJ, 804, 148
\bibitem[30]{} Wei, J.-J., Wu, X.-F. and Md Melia, F., 2014b, ApJ, 788, 190
\bibitem[31]{} Wei, J.-J., Wu, X.-F., Melia, F. and Maier, R. S., 2015a, AJ, 149, 102
\bibitem[32]{} Wei, J.-J., Wu, X.-Felia, F., 2013, ApJ, 772, 43
\bibitem[33]{} Wei, J.-J., Wu, X.-F., Melia, F., Wei, D.-M. and Feng, L.-L., 2014a, MNRAS, 439, 3329
\bibitem[34]{} Wei, J.-J., Wu, X.-F. an. and Melia, F., 2015b, MNRAS, 447, 479
\bibitem[35]{} Weigel, A. K., Schawinski, K., Treister, E., Urry, C. M. and Koss, M. and
Trakhtenbrot, B., 2015, MNRAS, 448, 3167
\bibitem[36]{} Willott, C. J., McLure, R. J. and Jarvis, M.J., 2003, ApJ Letters, 587, L15
\bibitem[37]{} Willott, C. J. et al., 2010, AJ, 140, 546
\bibitem[38]{} Wu, X.-B., et al. 2015, Nature, 518, Issue 7540, 512
\bibitem[39]{} Yoo, J. and Miralda-Escud\'e J., 2004, ApJ, 614, L25

\end{thebibliography}
\end{document}